\documentclass[twocolumn]{aastex631}
\usepackage{amsmath}
\usepackage{xcolor}
\usepackage[mathlines]{lineno}

\begin{document}
\title{Absorption of GRB X-ray Afterglows by The Missing Baryons: \\
Confronting Observations with Cosmological Simulations}

\author{Matan Grauer}
\author{Ehud Behar}
\affiliation{Department of Physics, Technion, Haifa 32000, Israel}


\begin{abstract}

A large fraction of the baryons at low redshift are undetected, and likely reside in the
tenuous, hot intergalactic medium (IGM).
One way to probe the missing baryons is through their absorption of bright sources. 
The anomalous absorption excess in the X-ray afterglows of $\gamma$-ray bursts (GRBs) has been suggested to result from the missing baryons.
In order to test this hypothesis, the present paper employs IllustrisTNG simulations to compute the X-ray absorption effect on cosmological distances.
The simulation shows that ionization of H and He in the IGM leaves the metals responsible for $>60\%$ of the X-ray opacity of high-$z$ sources 
The high-$z$ asymptotic optical depth at 0.5\,keV in the simulation reaches $0.15\pm0.07$, while the GRB afterglow values tend to $\approx 0.4$, implying the missing baryons can account for a significant fraction of the observed opacity.
The remaining discrepancy is ascribed mainly to the low average metallicity in the simulation, which drops from 0.06 solar at $z=0$ to 0.01 at $z=3$, and which is below previously measured values.
\end{abstract}

\section{INTRODUCTION}
Anomalous absorption of the X-ray afterglow of gamma-ray bursts (GRBs) is prevalent, and known since the days of BeppoSAX \citep{Stratta_2004}. 
Beyond the expected Galactic absorption there is an excess column density of $\sim 10^{21}$\,cm$^{-2} - 10^{23}$\,cm$^{-2}$ that remains to be explained.
This excess absorption is detected through attenuation of the soft X-rays that increase towards low energies, and is measured down to $\sim 0.3$\,keV.  
Such attenuation is typical of photo-electric (ionization) absorption.
Identifying the origin of this excess absorption is challenging, because the spectra are void of any discrete features, such as absorption lines or edges.
The common practice is to assume an absorber that is a) neutral b) with solar abundances, and c) at the redshift $z$ of the GRB, and to quote an equivalent column density $N_{\rm H}$. Lacking any supporting evidence in the data for these assumptions, one needs to keep in mind that the actual column density of the absorber depends strongly on all three of them, and may be significantly different.
For example, partial ionization, or sub-solar abundances will increase the derived $N_{\rm H}$. 

The X-ray telescope (XRT) on board the Swift observatory \citep{Gehrels_2004} considerably increased the sample size of GRB afterglow spectra. 
The equivalent $N_{\rm H}$ increases dramatically with $z$
from $\sim 10^{21}$\,cm$^{-2}$ at low-$z$ up to $\sim 10^{23}$\,cm$^{-2}$ at $z>5$ \citep{Campana_2010}. 
This result holds regardless of the Galactic absorption model \citep{Willingale_2013, Arcodia_2016}, the sample selection \citep{Campana_2012}, or the variability of the X-ray afterglow \citep{Valan_2023}.
\citet{Rahin_2019} later found the same trend with a larger sample. 
This trend with $z$ is curious as no other evolution of GRBs with redshift is known. 
Additionally, the equivalent X-ray $N_{\rm H}$ can be orders of magnitude higher than the Ly\,$\alpha$ column density readily measured at the GRB host \citep{Watson_2007, Campana_2010}, and which has no redshift evolution \citep{Rahin_2019}.

The directly measured quantity in GRB X-ray afterglows is not $N_{\rm H}$, but the optical depth $\tau = \sigma N_{\rm H}$. The fact that for $z \gtrsim 2, \tau(z)$ tends to a constant $\tau(0.5$\,keV)$\approx 0.4$, while $N_{\rm H}$ increases strongly with $z$, points to a varying absorption cross section ($\sigma$) with $z$, which is 
exactly expected from the photo-ionization cross section that decreases strongly with energy ($z$).
This physical effect led \citet{Behar_2011} to argue that the observed X-ray absorption may be due to the missing baryons in the intergalactic medium (IGM), while the host contribution to the 0.5\,keV opacity decreases with $z$, becoming negligible at $z>2$. 
Most of the IGM opacity accumulates at low-$z$ eventually saturating at $z>3$. 
Assuming a cosmologically uniform, neutral absorber, and an approximate cosmic cross section of $\sigma \propto E^{-2.5}$, they showed that the asymptotic X-ray optical depth at high-$z$ depends only on metallicity

\newpage

\begin{equation}
\tau_{\mathrm IGM} (z\rightarrow \infty)= \frac{2Z_0}{1+k}
\label{eq.:tau_IGM}
\end{equation}

\noindent where $Z_0$ is the metallicity at $z=0$ in solar units, and $k$ represents the metallicity evolution $Z(z)=Z_0(1+z)^{-k}$.

\citet{Starling13} further accounted for ionization in the IGM at $10^5 - 10^{6.5}$\,K and treated H and He separately from metals. They showed that the IGM could explain the observed X-ray opacity at $z > 3$, provided it is metal enriched to  $\approx 0.2$ solar metallicity. However, they confirmed that intervening cold gas could also account for the observed opacity.
\citet{Wang_2013} found a correlation between the $N_{\rm H}$ of GRB afterglows and that of IGM intervening systems.
\citet{Dalton_2021} confirmed the significant contribution of the IGM to $N_{\rm H}$. 
The IGM interpretation is further supported by the lack of absorption variability despite drastic changes in the X-ray afterglow flux \citep{Valan_2023}.
Alternative explanations for the discrepancy between $N_{\rm H}$ and host opacity have been suggested \citep{Watson_2011, Watson_2013, Krongold_2013, Tanga_2016, Heintz_2018}.
However, none of these explains the $z$ dependence of $N_{\rm H}$.

The above IGM hypothesis begs for confrontation with cosmological simulations.
\citep{Campana_2015} used the RAMSES code \citep{Teyssier_2002} to show that 20\% of the metals in the universe can reside in the IGM and around galaxies, and can explain the X-ray opacity towards distant X-ray sources. 
In this paper, we use the TNG-Illustris simulations \citep{Nelson_2018_2} to compute X-ray opacities at cosmic distances.
Illustris simulations include star formation, metal enrichment, and radiative feedback that were not available for \citet{Campana_2015}.
In the following sections we describe the simulations (Sec.\,\ref{sec:sim}), the computed opacities (Sec.\,\ref{sec:opacity}), the computational results (Sec.\,\ref{sec:comp_res}). We then compare these opacities and their $z$ dependence with Swift GRB $N_{\rm H}$ measurements in Sec.\,\ref{sec:comp}, and draw our conclusions in Sec.\,\ref{sec:conclusions}.


\section{Simulations}
\label{sec:sim}
We employ the publicly available IllustrisTNG simulation \citep{Nelson_2018_2, Nelson_2019_3, Pillepich_2019} to compute optical depths towards sources at different redshifts.
We worked with TNG50–4, which evolves a cube of 350 co-moving Mpc (cMpc$/h$) on the side.
It has a dark mass particle resolution of $2.8\times 10^8M_\odot$. 

Higher resolution and Larger simulations are available but require computing resources that are prohibitive for us.
TNG50–4 has 100 snapshots,
each snapshot captures a specific redshift.
We use 96 of them between $z = 0 - 10$.
Metallicity and neutral H fraction are available for only 18/96 snapshots. These 18 snapshots are distributed all along the redshift range $0 \leq z \leq 10$, in steps of approximately $\Delta z = 0.1$ for $0\leq z \leq 1$ and $\Delta z=1$ for $1<z\leq 10$.  \footnote
{https://www.tng-project.org/data/docs/specifications}

\begin{figure} [!hbtp]
    \centering
    \includegraphics[width = 0.2325\textwidth]{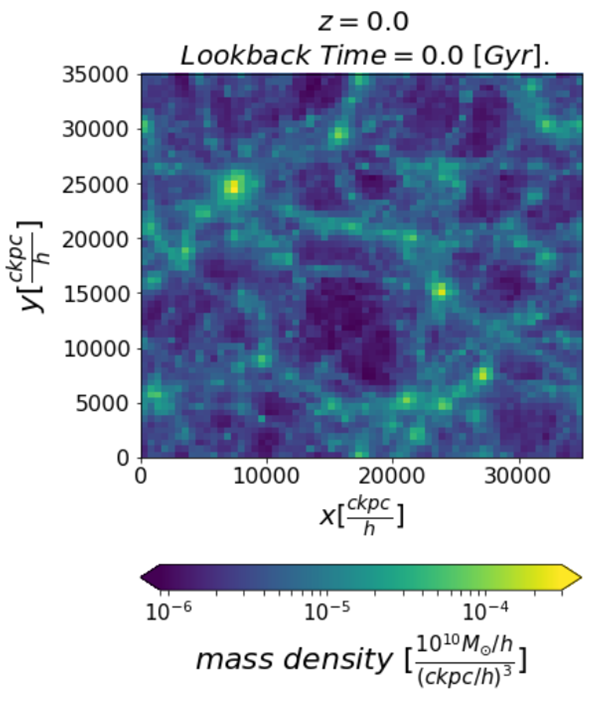}
    \includegraphics[width = 0.2325\textwidth]{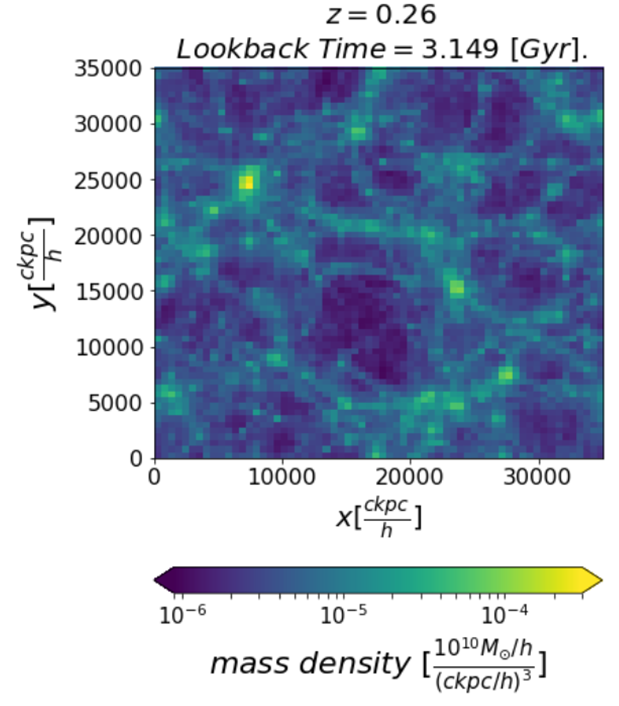}
    \includegraphics[width = 0.2325\textwidth]{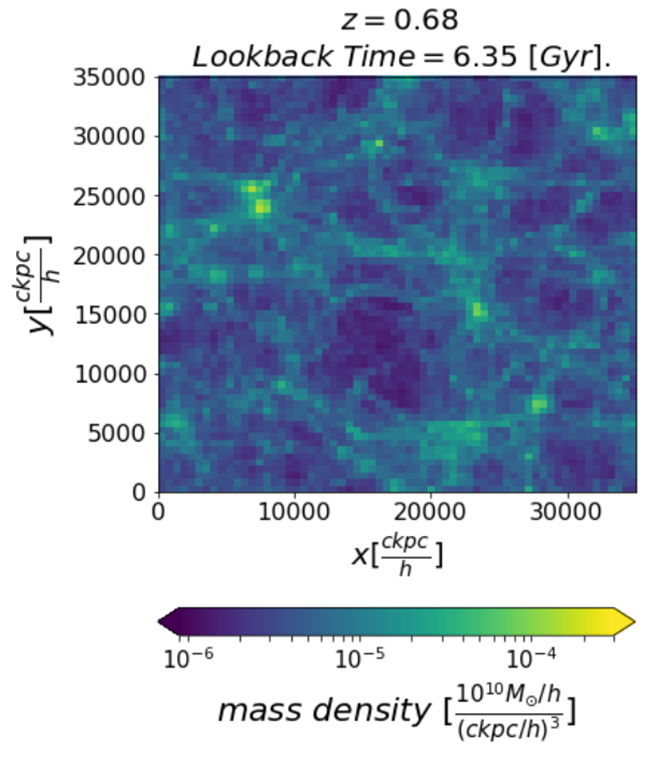}
    \includegraphics[width = 0.2325\textwidth]{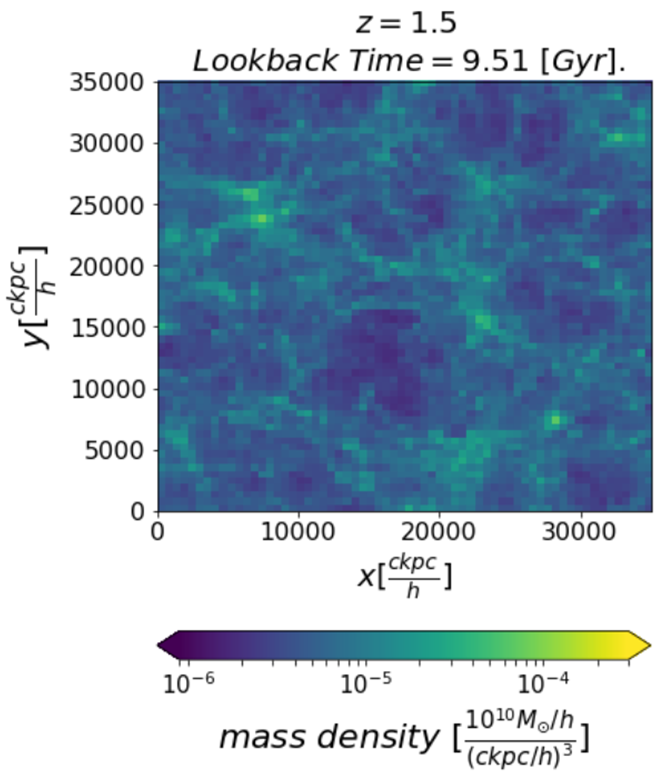}
    \includegraphics[width = 0.2325\textwidth]{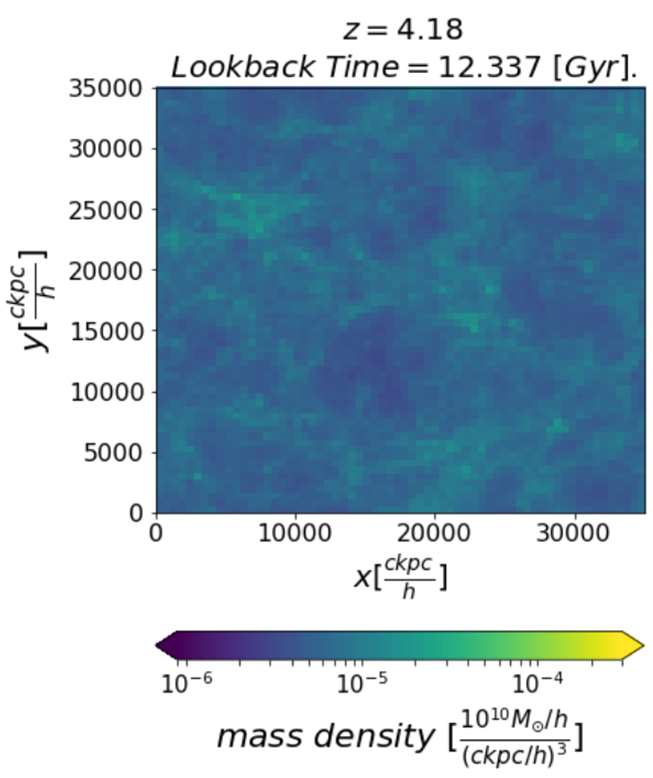}
    \includegraphics[width = 0.2325\textwidth]{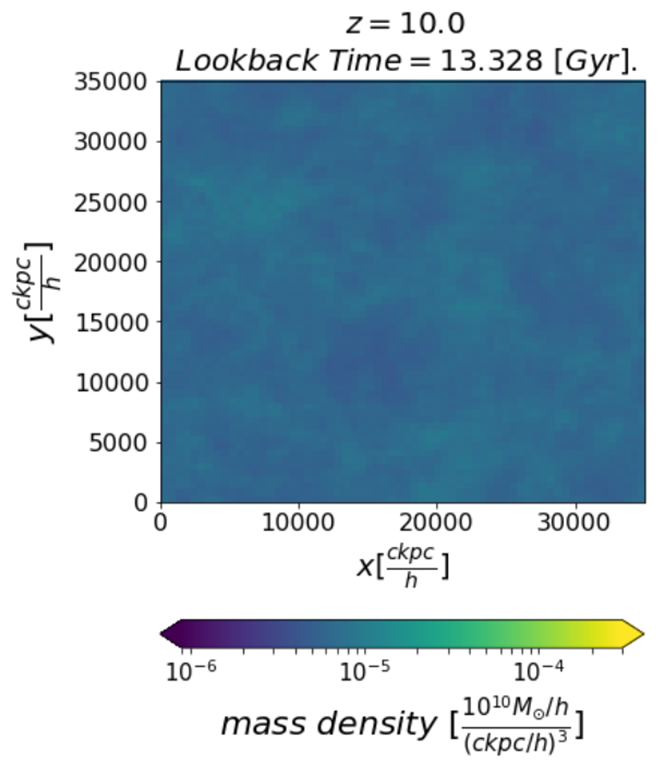}
    \caption{Simulated 2D-collapsed baryonic mass density maps on logarithm scale, binned to a $64\times64$ grid, demonstrating the formation of structure from $z = 0 - 10$.}   
    \label{fig:sim}
\end{figure}

In the other snapshots, we linearly interpolated. 
For column density measurements, we collapse each snapshot to two dimensions. 
Six such collapsed snapshots are shown in Fig.\,\ref{fig:sim} demonstrating structure formation from $z = 0 - 10$. 




\newpage

\section{Opacity computations}
\label{sec:opacity}

\begin{figure} [!htbp]
    \centering
    \includegraphics[width = 0.45\textwidth]{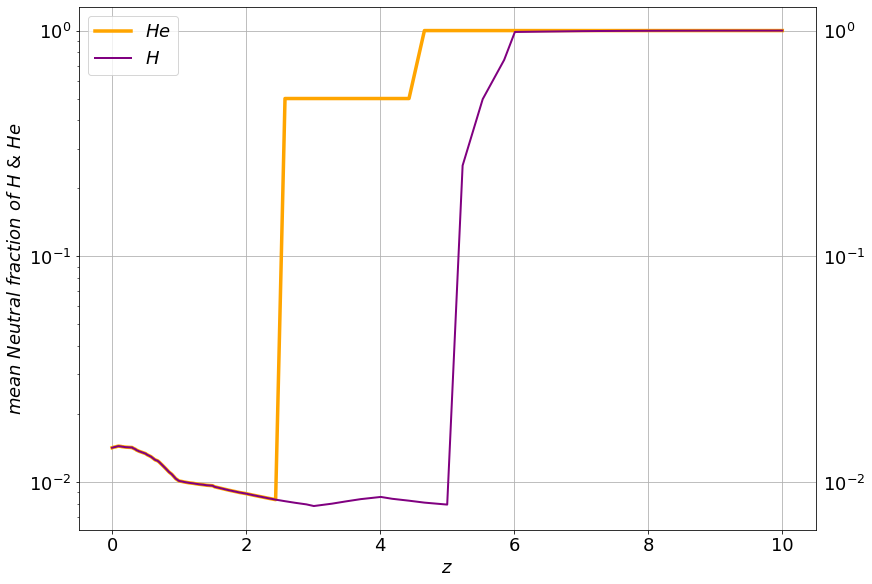}
\caption{Neutral fraction of H and He used in our computations. The H neutral fraction is taken from the simulations and the He one is assumed. }
\label{fig:ionization}
\end{figure}

In order to compute the column densities and optical depths, we need to extract the number density from the simulations, and integrate over cosmic sightlines.
To obtain the average mass density $\rho$ in each bin of the 2D collapsed ($64\times64$) grid (Fig.\,\ref{fig:sim}), we take its total mass and divide by its co-moving volume 350*(350/64)$^2$\,\,$\sim 10^5$(cMpc$/h)^3$. 
The physical number density of each species $s$ (e.g. H, He, ...) is then 

\begin{equation}
    n_s = \frac{\rho A_s(1+z)^3}{M_s}
\end{equation}

\noindent where $A_s$ is the average abundance of the $s$ species in the bin and $M_s$ is the mass of that species. The factor of $(1+z)^3$ gives the physical density at the snapshot of redshift $z$.
Fully ionized atoms do not contribute to the photo-electric optical depth. 
The ionization of H is readily given by the simulations, and features a sharp ionization phase at $z=5-6$ leaving only $\sim1\%$ of H neutral.
The He and metals ionization is not provided by the simulation.
We assume that He is ionized once at $z\approx4.5$ (its first ionization potential is higher than H), and is fully ionized at $z\approx2.5$. This is roughly where observations of IGM absorption systems indicate He is fully ionized  \citep{Dixon_2009, Furlanetto_2008, Sanderbeck_2020}.
At $z<2.5$, we assume the low neutral He fraction follows that of H from the simulation. 
The neutral fractions of H and He used in the present computations are summarized in Fig.\,\ref{fig:ionization}.
Metal ionization is neglected since partly ionized metals still absorb X-rays by K-shell (1s-electron) photo-ionization. Indeed, most of the IGM in IllustriusTNG is on small-scale
filaments dominated, e.g., by Li-like O, and lower charge states \citep{10.1093/mnras/sty656}.

\begin{figure} [!htbp]
    \centering
    \includegraphics[width = 0.45\textwidth]{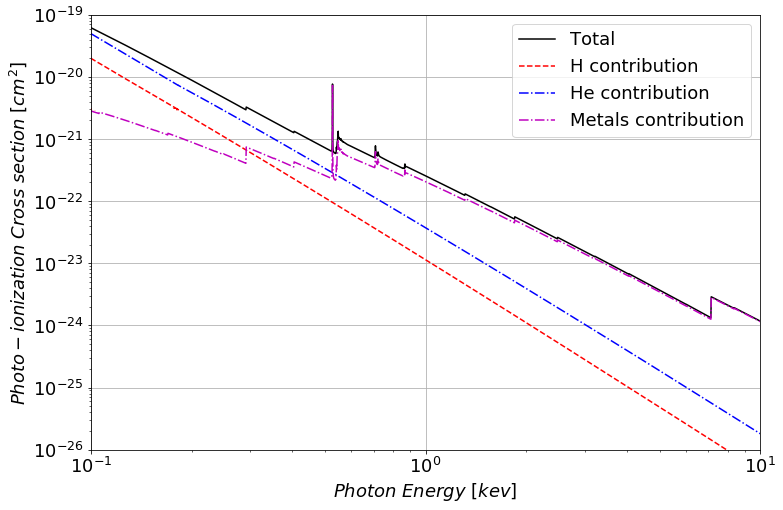}
    \label{fig:cs}
    \caption{Photoionization cross section per H atom extracted from the Xspec phabs model using the \citet{Wilms_2000} solar abundances, separated into its H, He, and metals contributions.}
\end{figure}

In order to compute the optical depth in the simulation, we sum separately over the neutral atomic species of H, He and all metals (Z) treated as a whole. 

\begin{equation}
\tau(E_0,z)=\sum_{s\in\left\{ H,He,Z\right\} }
\int_{0}^{z}n_s(z')\sigma_s[E(z')]c\frac{dt}{dz'}dz'
\label{eq:taushort}
\end{equation}

\noindent where $\sigma_s$ is the photo-electric cross section. 
 (insert*) For the metals, we use a cross section weighted by relative solar abundances. 
For absorption at $z', \sigma_s(E)$ needs to be evaluated at the de-redshifted energy of $E(z') = E_0(1+z')$.
Since $\sigma_s$ strongly decreases with energy, as can be seen in Fig.\,\ref{fig:cs}, higher-$z$ absorbers require higher (column) density to match the observed $\tau$. 
c is the speed of light. 

In terms of the cosmological model, $\tau$ is written as

\begin{multline}
        \tau (E_0,z)= \\
    \frac{c}{H_0}\sum_{s\in\left\{H,He,Z\right\}}
    \int_{0}^{z}\frac{n_s(z')
    \sigma_s[E_0(1+z')]}{(1+z')\sqrt{\Omega_M(1+z')^3+\Omega_\Lambda}}dz'
\label{eq:tau}
\end{multline}

\noindent where $H_0=67.7$\,km\,s$^{-1}$Mpc$^{-1}$ is the Hubble constant. $\Omega_M=0.31$ and $\Omega_\Lambda=0.69$ are the mass and dark-energy fractions of the present day energy density of the universe. 
The integral over $z'$ in Eq.\,\ref{eq:tau} is carried out as a discrete sum over the simulation snapshots, in $dz'$ steps determined by the $\Delta z'$ between snapshots. 
Random lines of sight are chosen in each snapshot, to avoid the line of sight going through the same structure over and over again (Fig.\,\ref{fig:sim}).
Structures larger than the simulation box, thus, can not be accounted for. 
We repeat the calculation of Eq.\,\ref{eq:tau} 300 times in order to study the statistics over many random lines of sight, similar to real GRB samples.
A consistency check of the simulation confirms an average co-moving baryon density of $2\times10^{-7}$\,cm$^{-3}$, and hydrogen column densities (computed by Eq.\,\ref{eq:taushort}, but without $\sigma$) of $N_{\rm H} = 4\times10^{21}$ up to $z = 1$ and $10^{23}$\,cm$^{-3}$ up to $z=10$, which is consistent with mean analytical values \citep[e.g.,][]{Behar_2011}.

\section{Computational Results}
\label{sec:comp_res}
\begin{figure} [!htbp]
    \includegraphics[width = 0.45\textwidth]{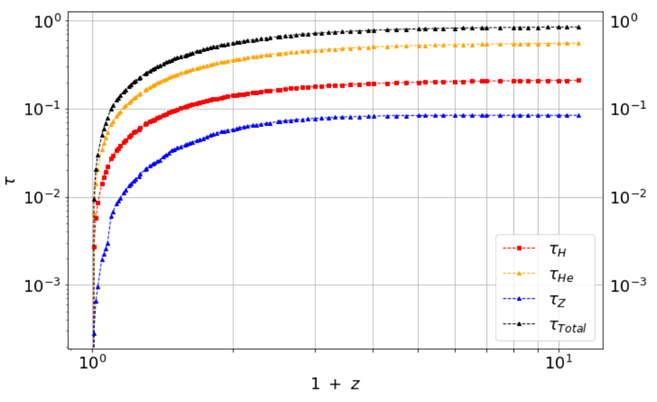}
    \includegraphics[width = 0.45\textwidth]{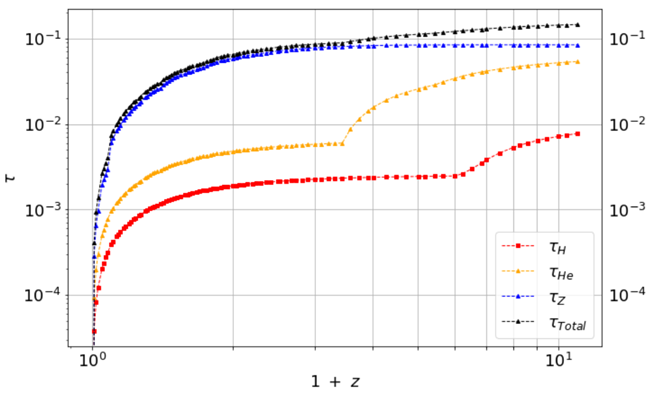}
    \caption{Optical depth $\tau(E=0.5$\,keV) averaged over 300 random lines of sight {\it Top}: assuming all atoms are neutral. {\it Bottom}: including ionization of H and He according to Fig.\,\ref{fig:ionization}.}
\label{fig:tau}
\end{figure}

Fig.\,\ref{fig:tau} shows the  results of our calculations for $\tau(z)$ according to Eq.\,\ref{eq:tau} and averaged over 300 random lines of sight.
The $\tau(z)$ values in random lines of sight represent the opacity of the IGM expected for random GRB locations.
We tested two cases.
In the first case, we assumed all atoms were neutral. The results are shown in the upper part of Fig.\,\ref{fig:tau}.
It can be seen that $\tau$ at high-$z$ approaches 0.9, which is higher than what the GRB afterglow observations indicate. Also, it can be seen that the He opacity dominates $\tau$ over H and the small contribution of metals.
When we include the effect of ionization (Fig.\,\ref{fig:ionization}) the contribution of H and He becomes negligible for $z<3$, while the metals dominate.
The values obtained at $z=10$ are $\tau_{\rm H} \approx 0.008, \tau_{\rm He} \approx 0.05$, and $\tau_{\rm Z} \approx 0.09$ for a total of $\tau_{\rm total} \approx 0.15$. This can be seen in the lower part of Fig.\,\ref{fig:tau}.
The figure shows that $\tau$ is not sensitive to the exact epoch in which He fully ionized, unless it ionized at $z < 2.5$, which could increase $\tau$.

\begin{figure}  [!hbp]
    \centering
    \includegraphics[width = 0.45\textwidth]{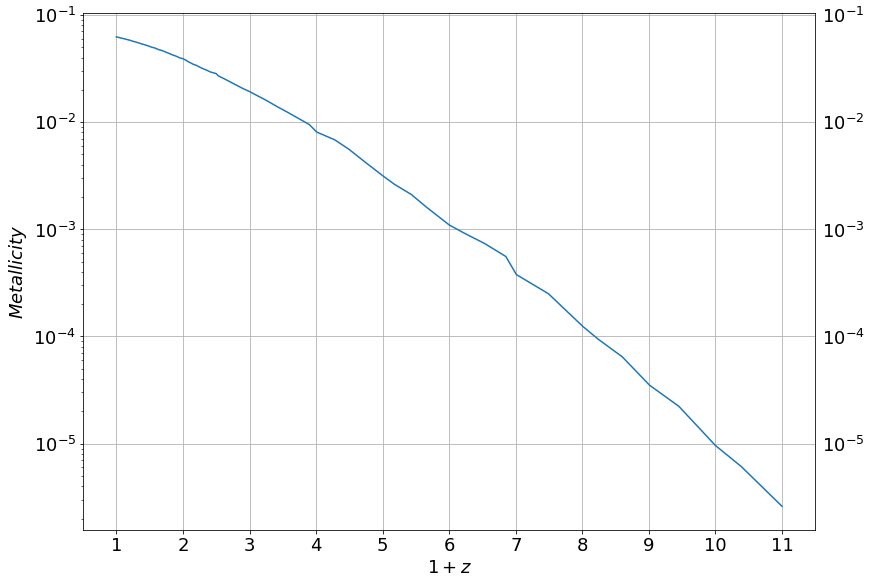}
    \caption{Average metallicity as a function of redshift from the simulation.}
    \label{fig:metals}
\end{figure}

The result above includes ionization, which is more physical and shows that $\tau$ is determined almost entirely by the metallicity, and mostly the metallicity at low-$z$.
Interestingly, $\tau \propto Z_0$ agrees well with the crude approximation of a homogeneous IGM presented in Eq.\,\ref{eq.:tau_IGM}.
The average metallicity, however, in the simulation is surprisingly low, peaking at $Z_0 \approx 0.06$ at $z=0$, and decreasing below 0.01 at $z=3$, as shown in Fig.\,\ref{fig:metals}. In terms of Eq.\,\ref{eq.:tau_IGM}, the metllicity index is $k=1.7$.
These metalicities are low compared to 
measurements of metallicity of $\sim 0.3$ (median) in circumgalactic media \citep{Prochaska_2017}, and a metallicity approaching solar in the IGM along lines of sight to GRBs and quasars \citep[see compilation in Fig.\,2 of][]{savaglio_2009_C}.


\begin{figure}  [!hbpt]
    \centering
    \includegraphics[width = 0.45\textwidth]{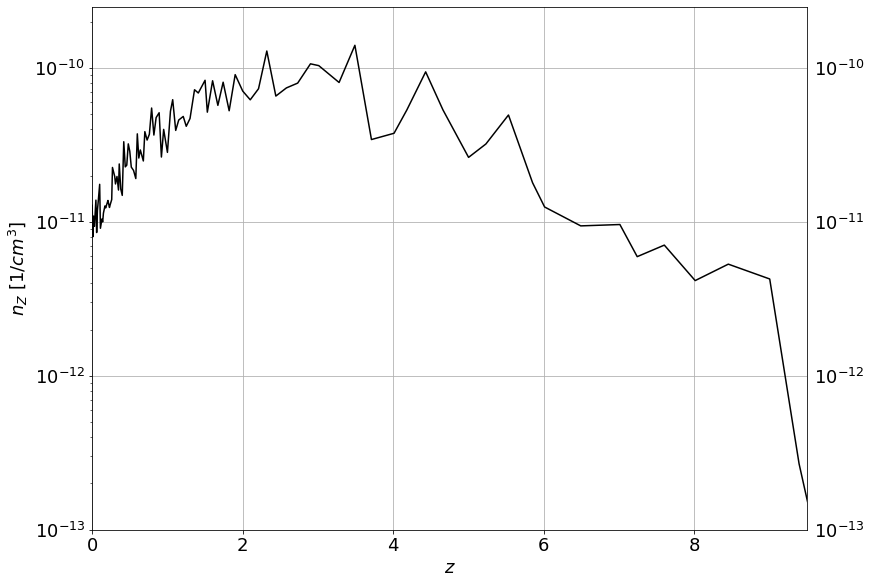}
   
    \caption{$\langle n_z \rangle$, mean metals number density as a function of redshift from the IllustrisTNG simulation.}
\label{fig:n_Z}
\end{figure}

The factor that goes into the calculation of $\tau$ in Eq.\,(\ref{eq:tau}) is not the metallicity, but the physical metal density $n_M(z')$. 
Fig.\,\ref{fig:n_Z} 
shows $n_M$ as a function of $z$ averaged over 300 lines of sight. $n_M$ increases with star formation, peaking between $z=2-3$, and decreases at low-$z$ due to the expansion of the universe. 
This behavior [along with the decreasing $\sigma_Z(E)$] explains why $\tau$ increases in Fig.\,\ref{fig:tau} up to $z\sim2$ and not beyond.
The dominance of metals in the bottom panel of Fig.\,\ref{fig:tau} deems Eq.\,\ref{eq.:tau_IGM} a  good approximation demonstrating that $\tau_{\rm IGM} (z \rightarrow \infty)$ is determined predominantly by the metallicity at low-$z$.

\newpage
\section{Comparison with GRB data}
\label{sec:comp} 
\begin{figure} [!htbp]
    \centering
    \includegraphics[width = 0.45\textwidth]{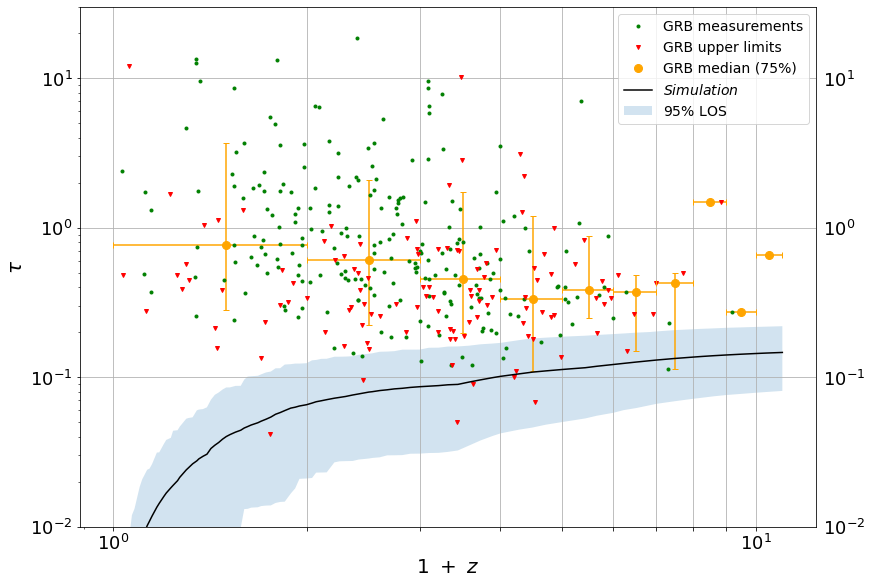}
    \caption{Optical depth $\tau(E=0.5$keV) of GRB X-ray afterglows. 
    The data from \citet{Rahin_2019} include 230 measured values (green points) and 123 upper limits (red triangles). Orange points are the medians over $\Delta z = 1$ bins (including upper limits) with error bars encompassing 75\% (6 central octiles) of the bursts.
    The average of 300 lines of sight (LOS) in the simulation is plotted as a solid curve, with the 95\% region shaded in blue.}
\label{fig:tau_sim_mes}
\end{figure}

The results of the IllustrisTNG simulation are compared in Fig.\,\ref{fig:tau_sim_mes} with 353 GRB Swift/XRT measurements of $\tau$, including 123 upper limits, taken from \citet{Rahin_2019}.
Overall, the measured GRB opacities are higher than the average values from the simulation. 
The two come close at high $z$; both data sets have significant scatter.
On average at high-$z$, the simulation results tend to $\tau = 0.15\pm0.07$ (95\%), while observed data tend to $\tau \approx 0.4$, starting at $z\sim 2-3$. Only a few GRBs are available for $z>5$.

The main reason for the low $\tau$ value in the simulation is its low average metallicity at low-$z$ (Figs.\,\ref{fig:metals}, \ref{fig:n_Z}). 
Complicating the comparison is the absorption in the host galaxy. 
We do not attempt to compute this contribution to $\tau$, but note that for a fixed host column density \citep[e.g., $N_{\rm H}= 10^{21}$\,cm$^{-2}$,][]{Behar_2011}, it will strongly diminish with $z$ , due to the decreasing cross section (Fig.\,\ref{fig:cs}).

Indeed, the difference between the measured $\tau$ values and those from the simulation decreases in Fig.\,\ref{fig:tau_sim_mes} with $z$.
However, the high-$z$ measured values remain about three times higher than in the simulation. 
In order for the IGM to explain the observed $\tau \approx 0.4$ at high-$z$, one needs a $z=0$ metallicity of $Z_0=0.2$, rather than $Z_0=0.06$ from the simulation.
Indeed, we compute $\tau$ with $Z_0=0.2$, and obtain at $z=10~\tau _{\rm Z} \approx 0.3$ and $ \tau _{\rm total} \approx 0.36$, within the measured range. 
Note that the median $\tau$ values in Fig.\,\ref{fig:tau_sim_mes} include the upper limits. 
Hence, the true median of $\tau(z)$ of GRB afterglows is below the plotted value, and closer to the simulation of the IGM.
This reduces the required host contribution. 






Another difference between the measured $\tau$ values and those from the simulation is the distribution around the mean. 
The 95\% range of lines of sight in the simulation is represented in Fig.\,\ref{fig:tau_sim_mes} by the blue shaded area. 
The distribution about the mean is rather normal and bounded by a factor of a few.
On the other hand, the distribution of observed values (green points and red triangles) is much broader and flatter, exceeding an order of magnitude in most redshifts.
At low-$z$ this can be ascribed to the variety of host galaxies and the GRB environment.
At $z>3$, where the IGM becomes important, the distribution narrows, yet suffers from a deficiency of measurements.

\section{CONCLUSIONS}
\label{sec:conclusions}

We employed the IllustrisTNG simulations to estimate the total IGM X-ray optical depth $\tau$ towards high-$z$ sources, and compare it with measured GRB afterglows.
The conclusions can be summarized as follows
\begin{itemize}
    \item Ionization effects in the IGM are important. Neglecting them results in an overestimation of $\tau$ by a factor of 6  (Fig.\,\ref{fig:tau}).
    
    \item The early ionization of H and He between $2.5 < z < 5$ (Fig.\,\ref{fig:ionization}) leaves the metals as the dominant ($>60\%$) X-ray absorbers in the IGM (Fig.\,\ref{fig:tau}, bottom). 
   
    \item The 0.5\,keV IGM opacity in the simulation is $\tau = 0.15\pm0.07$, which is sufficient to explain the observed values of $\approx 0.4$ (Fig.\,\ref{fig:tau_sim_mes}). 
    The opacity difference is explained by a residual host contribution, by the low metallicity in the simulation (Fig.\,\ref{fig:metals}), and by the numerous GRBs with $\tau$ upper limits (Fig.\,\ref{fig:tau_sim_mes}).
    
    \item The broad distribution of GRB-afterglow $\tau$ values at low-$z$ must be a result of the variety of host environments. At high-$z$, the distribution narrows, and better agrees with simulations. 
    (Fig.\,\ref{fig:tau_sim_mes}).
    
    \item The simulations show that the IGM can account for a major fraction of the missing baryons. 
    The exact contribution depends strongly on the IGM metallicity and ionization. These need to be better constrained by observations in order to benchmark the simulations.
\end{itemize}




This work was supported by a Center of Excellence of Israel Science Foundation (grant No. 1937/19). 
The authors thank Adi Nusser for useful discussions.
The IllustrisTNG simulations were undertaken with computer time awarded by the Gauss Centre for Supercomputing (GCS) under GCS Large-Scale Projects GCS-ILLU and GCS-DWAR on the GCS share of the supercomputer Hazel Hen at the High-Performance Computing Center Stuttgart (HLRS), as well as on the machines of the Max Planck Computing and Data Facility (MPCDF) in Garching, Germany.
We thank an anonymous referee for useful comments that improved the manuscript.

\newpage

\bibliography{sources.bib}
\bibliographystyle{aasjournal}


\end{document}